\documentclass[10pt,twocolumn,showpacs,preprintnumbers,amsmath,amssymb,aps,prl,superscriptaddress,longbibliography]{revtex4-1}
\usepackage{appendix}
\usepackage{bbm}
\usepackage{mathrsfs}
\usepackage{graphicx}
\usepackage{dcolumn}
\usepackage{bm}
\usepackage{braket}
\usepackage{amsmath}
\usepackage{amsfonts}
\usepackage[dvipsnames]{xcolor}
\usepackage[colorlinks=true,linkcolor=Blue,urlcolor=BlueViolet,citecolor=BlueViolet]{hyperref}
\usepackage{natbib}
\usepackage{floatrow}
\usepackage{multirow}

\begin{document}

\title{Anisotropic moir\'e band flattening in twisted bilayers of M-valley MXenes}
\author{Kejie Bao}
\thanks{These authors contribute equally to the work.}
\affiliation{State Key Laboratory of Surface Physics and Department of Physics, Fudan University, Shanghai 200433, China}
\affiliation{Shanghai Research Center for Quantum Sciences, Shanghai 201315, China}
\author{Huan Wang}
\thanks{These authors contribute equally to the work.}
\affiliation{State Key Laboratory of Surface Physics and Department of Physics, Fudan University, Shanghai 200433, China}
\affiliation{Shanghai Research Center for Quantum Sciences, Shanghai 201315, China}
\author{Zhaochen Liu}
\affiliation{State Key Laboratory of Surface Physics and Department of Physics, Fudan University, Shanghai 200433, China}
\affiliation{Shanghai Research Center for Quantum Sciences, Shanghai 201315, China}
\author{Jing Wang}
\thanks{Contact author: wjingphys@fudan.edu.cn}
\affiliation{State Key Laboratory of Surface Physics and Department of Physics, Fudan University, Shanghai 200433, China}
\affiliation{Shanghai Research Center for Quantum Sciences, Shanghai 201315, China}
\affiliation{Institute for Nanoelectronic Devices and Quantum Computing, Fudan University, Shanghai 200433, China}
\affiliation{Hefei National Laboratory, Hefei 230088, China}

\begin{abstract}
Experimental studies on moir\'e materials have predominantly focused on twisted hexagonal lattice with low-energy states near the $\Gamma$- or K-points, where the electronic dispersion is typically isotropic. In contrast, we introduce a class of semiconducting transition metal carbides (MXenes) $M_2$C$T_2$ ($M$ = Ti, Zr, Hf, Sc, Y; $T$ = O, F, Cl) as a new platform for M-valley moir\'e materials, which exhibit pronounced anisotropic properties. Using Ti$_2$CO$_2$ and Zr$_2$CO$_2$ as representative examples, we perform large-scale \emph{ab initio} calculations and demonstrate that their AB-stacked twisted homobilayer hosts three threefold rotational-symmetry-related M-valleys with time-reversal symmetry. These systems show striking anisotropic band flattening in the conduction band minimum. To elucidate the underlying physics, we construct a simplified moir\'e Hamiltonian that captures the essential features of the band structure, revealing the origins of anisotropic flattening through the mechanisms of band folding and interlayer tunneling. Our findings expand the current landscape of moir\'e materials, establishing valley- and spin-degenerate, two-dimensional arrays of quasi-one-dimensional systems as promising platforms for exploring many interesting correlated electronic phases.
\end{abstract}

\date{\today}

\maketitle

Moir\'e materials—synthetic two-dimensional crystals with large lattice constants arising from the twisting of layered 2D materials—have attracted substantial interest in recent years~\cite{Andrei2021,Kennes2021,balents2020,carr2020,Mak2022}. The flat electronic bands (relative to the scale of electron interactions) and highly tunable carrier filling in moir\'e heterostructures make them an exceptional platform for simulating a broad range of prototypical condensed matter systems. Prominent examples include twisted bilayer graphene~\cite{Bistritzer2011} and moir\'e superlattice based on transition metal dichalcogenides~\cite{Wu2018,Wu2019_tmd}, which exhibit a rich variety of interaction-driven phases, such as superconductors~\cite{Cao2018_2,yankowitz2019,lian2019,Xia2024,Guo2025}, 
correlated states~\cite{Cao2018_1,lu2019,Song2022,Kerelsky2019,Chou2023,Zhu2020,Tang2020,Claassen2022,Sheng2024,Wang2020,Anderson2023}, 
Chern insulators~\cite{Sharpe2019,Serlin2020,Chen2020,nuckolls2020,Wu2020,Dong2024,Tarnopolsky2019,Li2021,Devakul2021,Angeli2021,liu2022,Benjamin2024,Zhang2024,lian2020}, 
fractional Chern insulators~\cite{Lu2024,Jonah2024,Xu2023,Park2023,Zeng2023,Cai2023,Wang2024,Jia2024,Yu2024,Li2024} 
and so on~\cite{Saito2021,Huang2023,Yoo2019,Halbertal2021}.

Experimental studies on moir\'e materials have predominantly focused on twisted hexagonal lattices with the low-energy states near the $\Gamma$- or K-points. In these cases, the low-energy dispersion is typically isotropic due to the three-fold rotation symmetry $C_{3z}$. Consequently, when band flattening occurs in moir\'e patterns, the resulting bands remain nearly dispersionless throughout the entire moir\'e Brillouin zone (mBZ). This naturally raises the intriguing question: What occurs when the $C_{3z}$ symmetry is significantly broken? Twisted bilayer WTe$_2$ provides a compelling example, where pronounced transport anisotropy and power-law scaling of cross-wire conductance suggest the emergence of a sliding Luttinger liquid upon twisting~\cite{Wang2022}. These observations demonstrate that anisotropic band flattening—beyond the conventional isotropic scenario—is indeed possible in moir\'e systems~\cite{Kariyado2019,Rubio2020,fujimoto2022,Wang2023,Hu2024}, particularly in low-symmetry configurations. The significance of anisotropic band flattening in 2D moir\'e materials lies in its connection to the higher-dimensional generalization of the Luttinger liquid and to non-Fermi liquid behavior~\cite{Anderson1991,Lubensky2000,Lubensky2001,Vishwanath2001,sondhi2001,wu2024}. These theoretical frameworks have long been associated with enigmatic phases in condensed matter physics, including the normal state of cuprate superconductors~\cite{Anderson1990,Anderson1991,Wuyts1993}, quantum criticality~\cite{Classen2018,Lake2005} and unconventional metals~\cite{Jiang2013,Assa2018}. These insights motivate a closer examination of the underexplored M-point~\cite{Kariyado2019,Kariyado2023} in the hexagonal lattice, where symmetry breaking could enable new regimes of correlated electronic behavior.

In this Letter, we introduce a class of semiconducting MXenes—denoted as $M_2$C$T_2$ (where $M$ = Ti, Zr, Hf, Sc, Y; $T$ = O, F, Cl), as listed in Table~\ref{tab1}—as promising candidates for realizing M-valley moir\'e materials. Based on comprehensive \emph{ab initio} calculations, we demonstrate that AB-stacked twisted $M_2$C$T_2$ bilayer can exhibit anisotropic band flattening~\cite{supple}. To gain deeper insight, we construct a simplified moir\'e Hamiltonian for these systems and perform a detailed band structure analysis. Our results reveal that the M-valley moir\'e Hamiltonians possess an emergent spectral periodicity, protected by an effective symmetry that exchanges the M valleys of the top and bottom layers. This symmetry is preserved within a lowest-harmonic approximation of the model. We propose a modified mBZ that reflects this emergent symmetry and allows for a clear understanding of the mechanisms driving anisotropic band flattening—specifically, band folding and gap-opening processes.

\begin{figure}[t]
  \begin{center}
    \includegraphics[width=3.4in,clip=true]{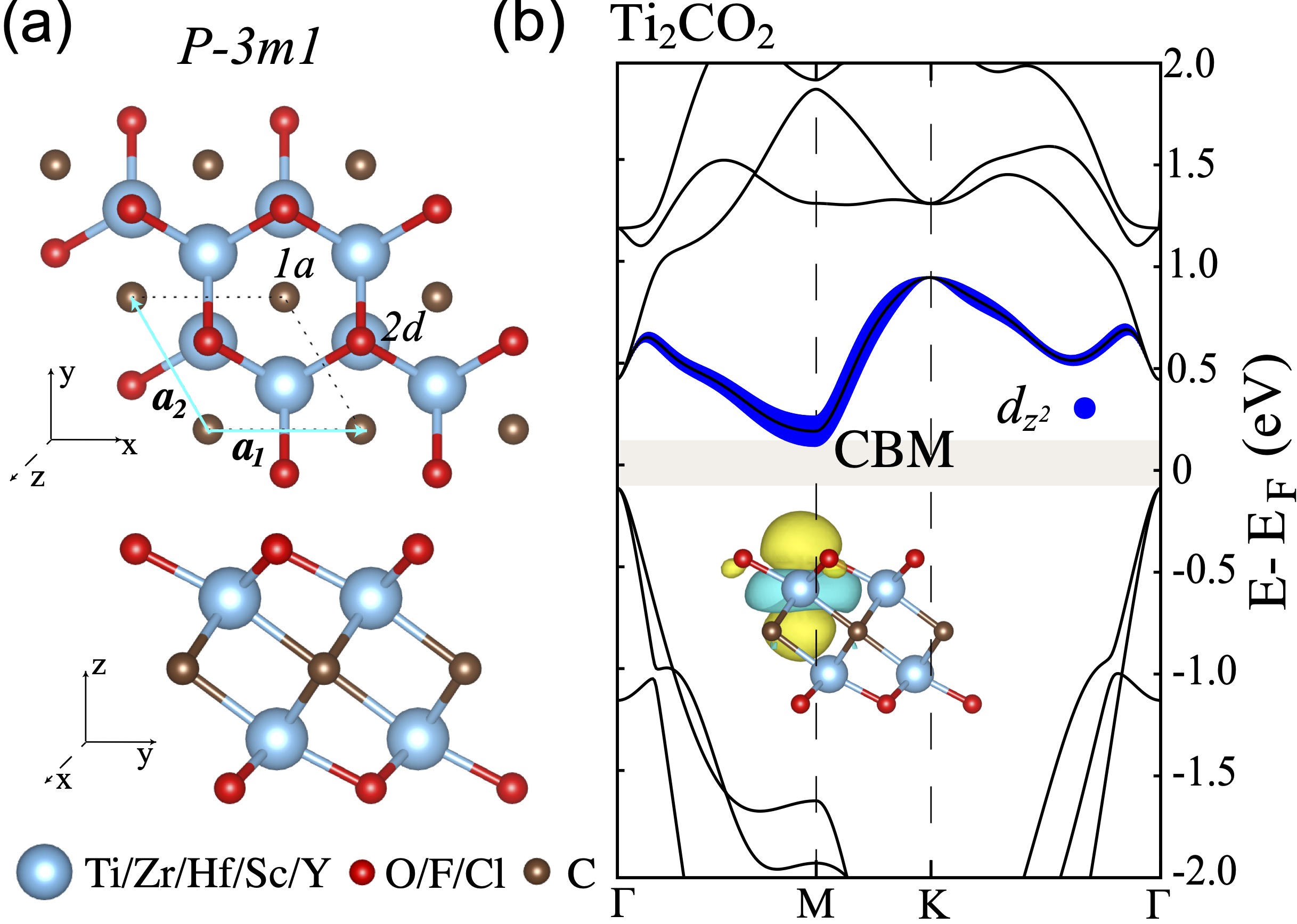}
      \caption{(a) Structures for monolayer MXenes $M_2$C$T_2$ ($M$ = Ti, Zr, Hf, Sc, Y; $T$ = O, F, Cl). The Wyckoff positions 1$a$ and 2$d$ are displayed (notation adopted from BilbaoCrystallographic Server). (b)  Band structure and band projections of $d_{z^2}$ orbital for monolayer Ti$_2$CO$_2$. The inner figure shows the Wannier functions that predominantly contribute to CBM at the M-valley.}
      \label{fig1}
  \end{center}
\end{figure} 

\emph{Materials.} Monolayer MXenes—transition metal carbides of the form $M_2$C$T_2$—crystallize in a triangular lattice with space group $P$-$3m1$ (No.~164), which includes inversion symmetry $\mathcal{I}$, out-of-plane three-fold rotational symmetry $C_{3z}$, and in-plane two-fold rotational symmetry $C_{2x}$. As shown in Fig.~\ref{fig1}(a), a central carbon layer is sandwiched between two layers of transition metal cations, while the anions are aligned directly above and below the cation layer on both side. Because the $d$ electrons of the $M$ atoms are transferred to the $p$ orbitals of the $T$ atoms, the ground states of these materials are spin-unpolarized, and thus preserve time-reversal symmetry $\mathcal{T}$. The corresponding lattice constants are listed in Table~\ref{tab1}. Remarkably, these materials have already been synthesized in experiments~\cite{naguib2012two,lai2015surface,melchior2018high,wang2023direct,druffel2019synthesis,hwu1986synthesis}, and have attracted growing interest due to their tunable band gaps~\cite{zhang2019prediction}, high carrier mobilities~\cite{zhang2015high}, and strong excitonic effect~\cite{dong2023robust}. Taking Ti$_2$CO$_2$ as a representative example, its band structure is shown in Fig.~\ref{fig1}(b). Notably, the conduction band minimum (CBM) resides at the M-point. The Wannier functions that contribute predominantly to the CBM are associated with the Ti $d_{z^2}$ orbitals. Due to the reduced symmetry at the M-point (retaining only $C_{2x}$), the effective mass at the M-valley is anisotropic along the $x$ (M-K) and $y$ (M-$\Gamma$) directions, as detailed in Table~\ref{tab1}.

\begin{table}[b]
\caption{Lattice constant, band gap, effective mass along the $x$ and $y$ directions for monolayer MXenes based on GGA method~\cite{perdew1996generalized}. The effective mass is in the unit of the free electron mass $m_0$.}
\begin{center}\label{tab1}
\renewcommand{\arraystretch}{1.3}
\begin{tabular*}{\columnwidth}
{@{\extracolsep{\fill}}ccccc}
\hline
\hline
 Materials& $a$ (\AA) &$E_g$ (eV) & $m_x$ ($m_0$) & $m_y$ ($m_0$) \\
 \hline
 Ti$_2$CO$_2$ & 3.04 &0.24    & 0.46 & 4.54\\
Zr$_2$CO$_2$ & 3.32 &0.99    & 0.35 & 2.74\\
Hf$_2$CO$_2$ & 3.27 & 1.03   & 0.28 & 2.15 \\
Sc$_2$CF$_2$ & 3.26 & 1.00  &  0.31 & 1.60\\
Y$_2$CF$_2$ & 3.66 &1.29    & 0.31 & 1.29\\
Sc$_2$CCl$_2$ & 3.43 &0.86   & 0.27 & 1.40\\
Y$_2$CCl$_2$ & 3.70 &0.94    & 0.23 & 1.07\\
\hline
\hline
\end{tabular*}
\end{center}
\end{table}

\begin{figure}[t]
  \begin{center}
    \includegraphics[width=3.4in,clip=true]{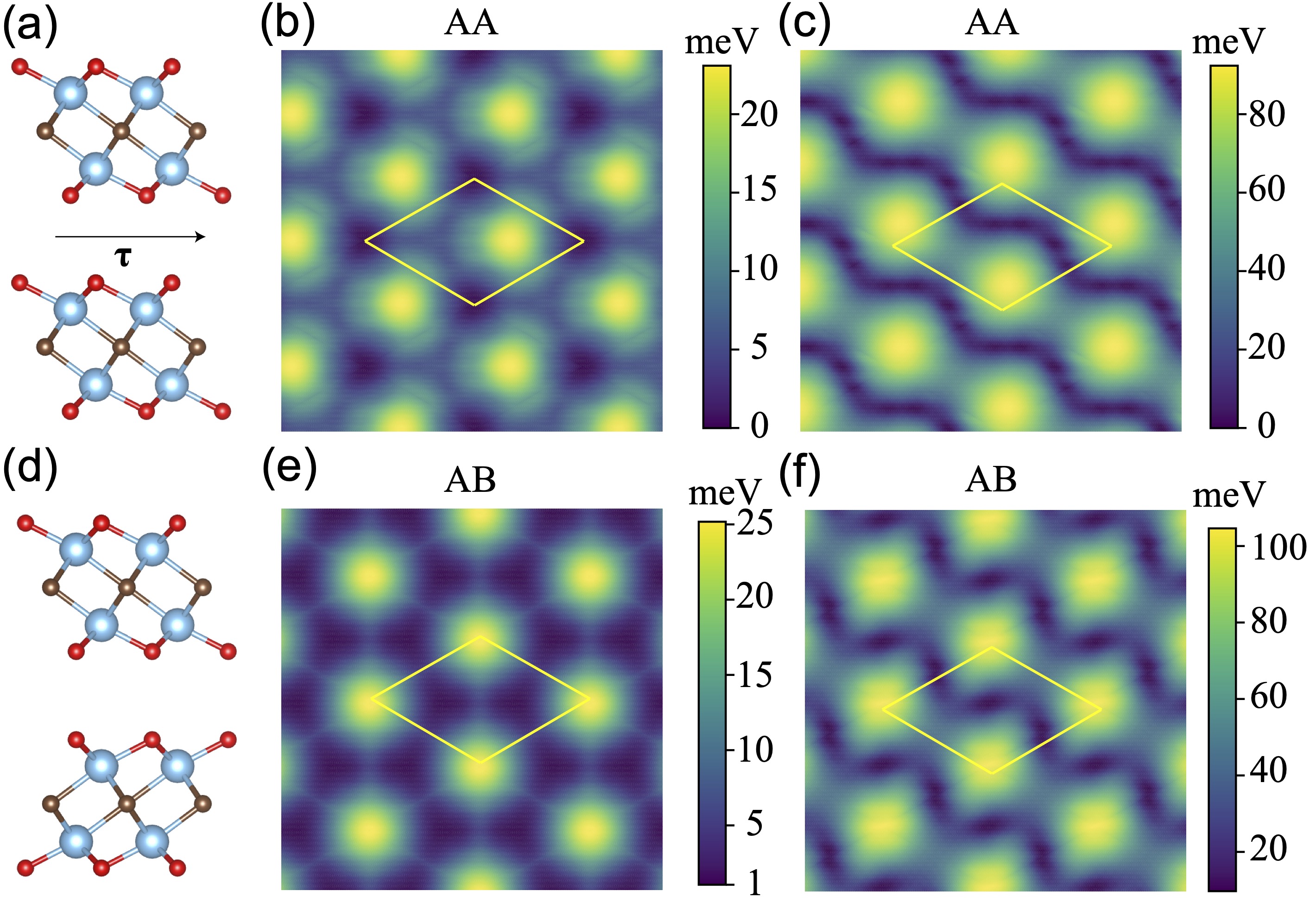}
      \caption{(a,d): Structures of AA and AB stacking for Ti$_2$CO$_2$ with in-plane displacement $\bm{\tau} = 0$. (b,e) Interlayer potential energy distribution for AA and AB stacking. (c,f) Energy splitting of M-valley at CBM for AA and AB stacking. The unit cell is illustrated by yellow lines.}
      \label{fig2}
  \end{center}
\end{figure}

We then investigated the interlayer tunneling of the untwisted Ti$_2$CO$_2$ bilayer. Due to the absence of out-of-plane rotational symmetry $C_{2z}$ in the monolayer, two distinct stacking configurations are considered: AA stacking, where the top and bottom layers are aligned directly on top of each other, and AB stacking, where the bottom layer is rotated by 180$^{\circ}$ prior to stacking, as illustrated in Fig.~\ref{fig2}(a) and Fig.~\ref{fig2}(d), respectively. Both configurations preserve the symmetries $\mathcal{I}$, $\mathcal{T}$, and $C_{3z}$, but differ in their in-plane rotational symmetries: AA stacking retains $C_{2x}$, while AB stacking exhibits $C_{2y}$. The interlayer potential energy distributions for each configuration are shown in Fig.~\ref{fig2}(b) and Fig.~\ref{fig2}(e). In the AA stacking, regions of high potential energy (indicated by brighter areas) are more densely distributed compared to the AB stacking, suggesting that lattice relaxation effects may be more significant in the AA-stacked twisted bilayer. To quantify interlayer tunneling, we analyze the M-valley band splitting for both configurations, shown in Fig.~\ref{fig2}(c) and Fig.~\ref{fig2}(f). The similar splitting patterns observed in both stackings indicate that their interlayer tunneling strengths are nearly equivalent.

\begin{figure}[t]
  \begin{center}
    \includegraphics[width=3.4in,clip=true]{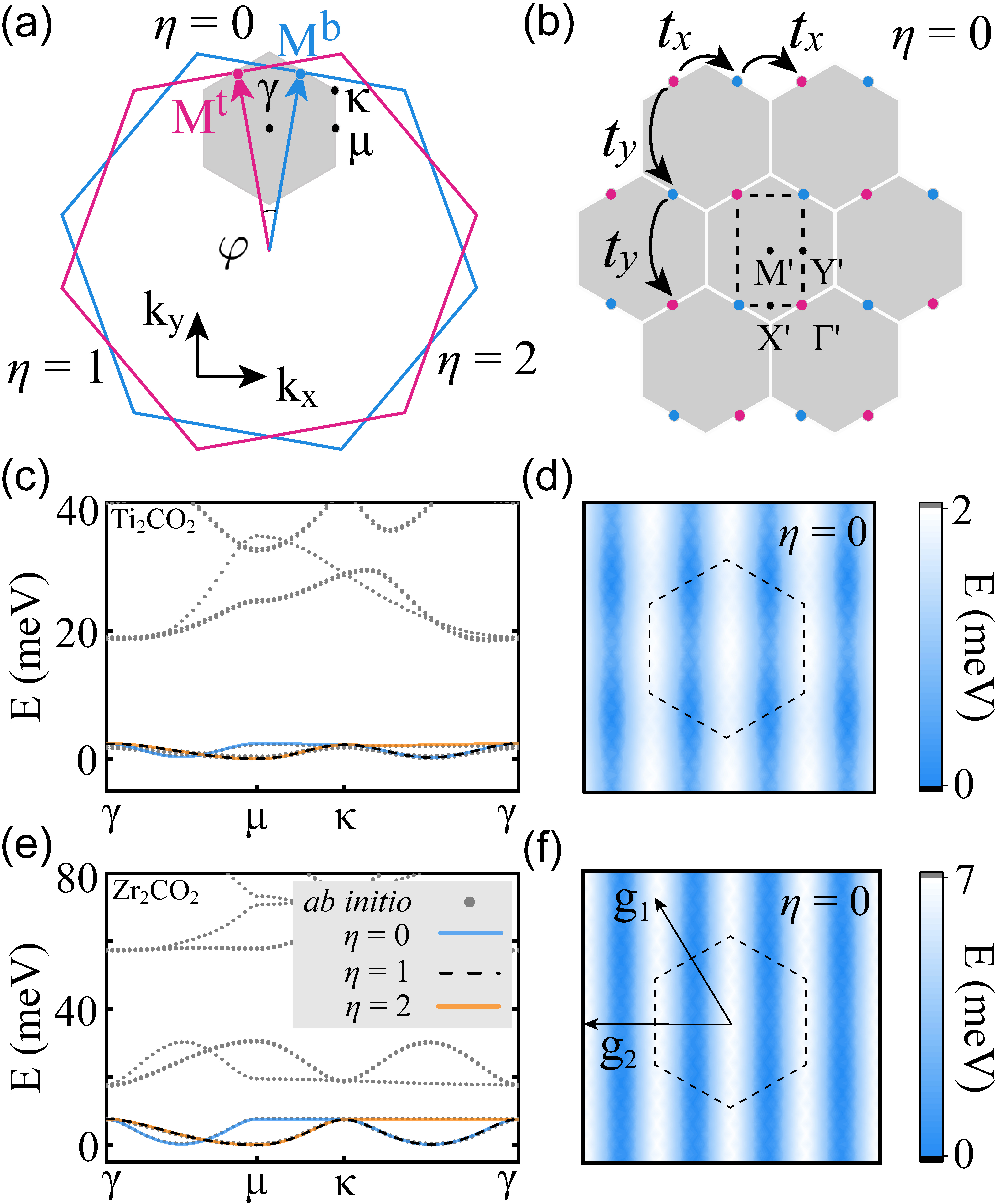}
      \caption{(a) Monolayer BZ and moir\'e BZ, represented as colorful and gray hexagons, respectively. Here M$^{t/b}$ labels the $\eta=0$ M-valley in top/bottom layer, where the momentum ${\mathbf{M}^t}=\mathbf{g}_1/2$, ${\mathbf{M}^b}=\left(\mathbf{g}_1-\mathbf{g}_2\right)/2$ in the mBZ. (b) The momentum space tight-binding model described by moir\'e Hamiltonian $\mathcal{H}_{\eta=0}$. Gray hexagons label the hexgonal mBZ and the dashed rectangle label the rectangle mBZ. (c,e) The band structure of twisted bilayer Ti$_2$CO$_2$ and Zr$_2$CO$_2$ at the twist angle $\varphi=3.89^\circ$, respectively. The results from density-functional theory are plotted as gray dot, and the fitting bands from moir\'e Hamiltonian $\mathcal{H}_{\eta}$ are plotted as different color and line style according to their valley index $\eta$. (d,f) The energy spectrum of the lowest band in moir\'e Hamiltonian $\mathcal{H}_{\eta=0}$ with fitting parameters $t_x=11$~meV, $t_y=24$~meV for Ti$_2$CO$_2$ and $t_x=20$~meV, $t_y=10$~meV for Zr$_2$CO$_2$.}
      \label{fig3}
  \end{center}
\end{figure}

\emph{M-vally moir\'e band.} We perform large-scale \emph{ab initio} calculations for twisted bilayer MXenes at a twist angle $\varphi=3.89^\circ$. The moir\'e lattice vectors are fixed as $\mathbf{L}_i=\mathbf{a}_i\times\hat{\mathbf{z}}/(2\sin(\varphi/2))$, and full atomic relaxation is allowed~\cite{supple}. In the relaxed structures, the symmetries $C_{3z}$, $\mathcal{T}$, and $C_{2x}$ ($C_{2y}$) are preserved for the twisted AA (AB) bilayers, while $\mathcal{I}$ is broken. Fig.~\ref{fig3}(c) and Fig.~\ref{fig3}(e) show the band structures of AB-stacked twisted bilayer Ti$_2$CO$_2$ and Zr$_2$CO$_2$, respectively. In both materials, the CBM consists of three spin-degenerate bands that are well-isolated from higher-energy bands. Among them, one band exhibits clear flatness along the M-K and K-$\Gamma$ directions, while retaining noticeable dispersion along the $\Gamma$–M direction. The corresponding bandwidths of the lowest-energy bands are approximately $2.2$~meV for Ti$_2$CO$_2$ and $7.6$~meV for Zr$_2$CO$_2$.

To elucidate the band structure observed in our \emph{ab initio} calculations, it is essential to construct a moir\'e Hamiltonian for the M-valley in twisted bilayer systems. We begin by analyzing the electronic structure of the aligned bilayer. Since the three M-valleys are related by $C_{3z}$, we initially focus on the $\eta = 0$ M-valley, as illustrated in Fig.~\ref{fig3}(a). The moir\'e Hamiltonians for the other two M-valleys, labeled $\eta = 1, 2$, can be generated by applying the $C_{3z}^\eta$ symmetry operator to the $\eta = 0$ case. In the AB-stacked MXene homobilayer, the CBM at M-valley is predominantly characterized by $d_{z^2}$ orbitals. These states approximately respect spin $SU(2)$ symmetry due to the negligible spin-orbit coupling. Based on this understanding, we construct the moir\'e Hamiltonian for the $\eta = 0$ M-valley as
\begin{equation}\label{eq1}
  \mathcal{H}_{\eta=0}=\begin{pmatrix} \frac{\left(k_x-\text{M}^t_x\right)^2}{2m_x}+\frac{\left(k_y-\text{M}^t_y\right)^2}{2m_y} & \Delta_T(\mathbf{r}) \\ \Delta_T^\dagger(\mathbf{r}) & \frac{\left(k_x-\text{M}^b_x\right)^2}{2m_x}+\frac{\left(k_y-\text{M}^b_y\right)^2}{2m_y} \end{pmatrix},
\end{equation}
where the diagonal terms describe the anisotropic kinetic energies centered at the $\eta = 0$ M-point momenta $\mathbf{M}^{t/b}$ in the top and bottom layers, respectively. The off-diagonal term $\Delta_T(\mathbf{r})$ represents the interlayer tunneling, which varies periodically with the moiré lattice vectors $\mathbf{L}_1$ and $\mathbf{L}_2$. By invoking the $C_{2y}$ and $\mathcal{T}$ symmetries of the M-valley and applying the two-center approximation~\cite{Bistritzer2011,jung2014}, the tunneling term takes the following lowest-harmonic form:
\begin{equation}\label{eq2}
  \Delta_T(\mathbf{r})=\left[t_x(1+e^{-i\mathbf{g}_2\cdot\mathbf{r}})+t_y(e^{-i\mathbf{g}_1\cdot\mathbf{r}}+e^{-i\left(\mathbf{g}_2-\mathbf{g}_1\right)\cdot\mathbf{r}})\right]\sigma_0,
\end{equation}
where $t_x$ and $t_y$ are real parameters characterizing the anisotropic tunneling amplitudes, $\mathbf{g}_i$ are the moir\'e reciprocal lattice vectors satisfying $\mathbf{g}_i \cdot \mathbf{L}_j = 2\pi\delta_{ij}$, and $\sigma_0$ is the $2\times2$ identity matrix acting on the spin degree of freedom. The spin-orbit coupling is neglected in the interlayer tunneling, consistent with the weak spin-orbit coupling in these materials.

Despite its simplicity—with only two fitting parameters—our moir\'e Hamiltonian effectively captures the essential low-energy physics of the M-valley moir\'e system. As shown in Fig.~\ref{fig3}(c,e), using parameters $t_x = 11$~meV and $t_y = 24$~meV for Ti$_2$CO$_2$, and $t_x = 20$~meV and $t_y = 10$~meV for Zr$_2$CO$_2$, the model accurately reproduces the lowest six conduction bands of the twisted AB-stacked structures. The individual contributions from the three symmetry-related M-valleys are distinguished by different colors and line styles, providing a clear decomposition of valley-resolved band features. To further understand the emergent low-energy behavior, we examined the energy spectrum of the lowest band from the $\eta = 0$ M-valley. Fig.~\ref{fig3}(d,f) reveal a pronounced quasi-1D character: the band is highly dispersive along the $x$-direction, while remaining nearly flat along the $y$-direction. This anisotropic dispersion underscores the directional nature of electronic states in the M-valley moir\'e system and is a hallmark of the anisotropic band flattening facilitated by the underlying lattice symmetry and interlayer coupling.

Upon closer examination, we find that the energy spectrum $E_\mathbf{k}$ exhibits a striking periodicity along the vectors $\mathbf{g}_2/2=\mathbf{M}^t-\mathbf{M}^b$ and $\left(\mathbf{g}_2-2\mathbf{g}_1\right)/2=-\mathbf{M}^t-\mathbf{M}^b$, as clearly shown in Fig.~\ref{fig3}(d). This periodicity arises from an emergent symmetry intrinsic to the simplified moir\'e Hamiltonian $\mathcal{H}_{\eta=0}$. The Hamiltonian functions as a tight-binding model in momentum space, where the hopping integrals between momentum states $\mathbf{M}^{t}$ and $\mathbf{M}^{b}$ along the $x$ and $y$ directions is governed by $t_x$ and $t_y$, respectively. As illustrated in Fig.~\ref{fig3}(b), $\mathcal{H}_{\eta=0}$ remains invariant under the exchange of $\mathbf{M}^t$ and $\mathbf{M}^b$, combined with a momentum space translation by $\mathbf{g}_2/2$ or $(\mathbf{g}_2-2\mathbf{g}_1)/2$. This invariance leads directly to the observed spectral symmetry: $E_\mathbf{k} = E_{\mathbf{k}+\mathbf{g}_2/2} = E_{\mathbf{k}+(\mathbf{g}_2 - 2\mathbf{g}_1)/2}$. This emergent symmetry is analogous to the effective $\tilde{M}_z$ symmetry described in Ref.~\cite{Calugaru2024}, and plays a pivotal role in shaping the electronic structure of M-valley moir\'e systems. It highlights the power of the minimal Hamiltonian in capturing key features of the system's low-energy physics. Importantly, this emergent symmetry is generally broken when higher-order harmonic terms are included. While such terms are negligible in twisted AB-stacked Ti$_2$CO$_2$ and Zr$_2$CO$_2$, they become significant in their AA-stacked counterparts~\cite{supple}.

\emph{Anisotropic band flattening at M-valley.} Given that the emergent periodicity in the energy spectrum $E_\mathbf{k}$ originates from an effective symmetry—namely, the exchange of M-valleys between layers—the conventional hexagonal mBZ becomes redundant for analyzing the band structure near the M-point. Instead, we adopt a rectangular mBZ that more accurately reflects the reduced symmetry of the M-valley moir\'e system. The reciprocal lattice vectors of this rectangular mBZ are chosen as $\mathbf{g}_2/2$ and $(\mathbf{g}_2-2\mathbf{g}_1)/2$, directly corresponding to the observed spectral periodicities. As shown by the dashed lines in Fig.~\ref{fig3}(b), this modified mBZ explicitly incorporates the emergent symmetry and provides a more natural framework for interpreting the anisotropic and quasi-1D features of the band structure.

\begin{figure}[t]
  \begin{center}
    \includegraphics[width=3.4in,clip=true]{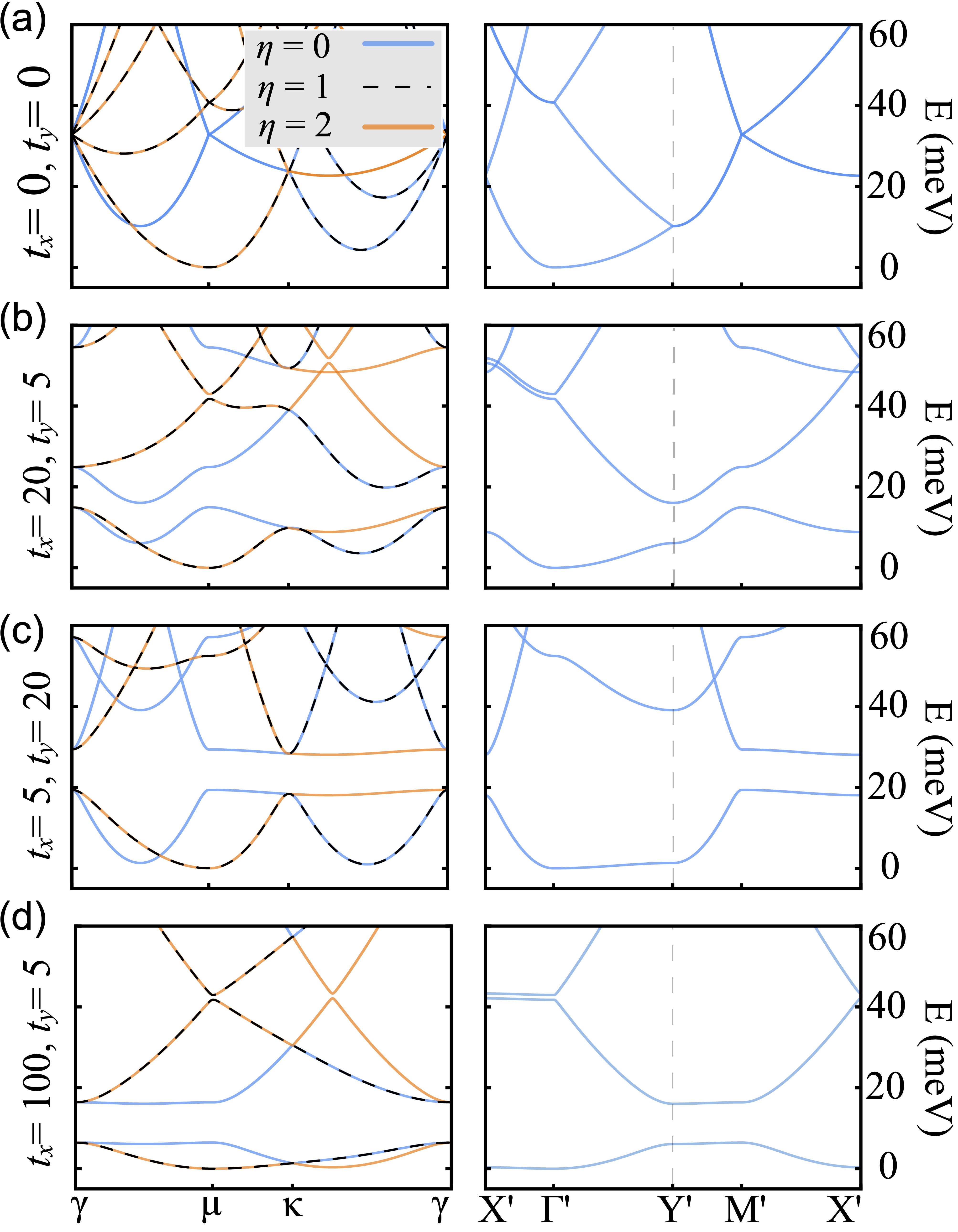}
      \caption{The band structure from moir\'e Hamiltonian Eq.~(\ref{eq1}) with typical $t_x$,$t_y$ parameters plotted in the hexagonal (left column) and rectangular mBZ (right column). The different line color and style label the contribution from different M-valleys in hexagonal mBZ, while in rectangular mBZ only the $\eta=0$ M-valley is plotted. Here we set $\varphi$ = 4$^{\circ}$, lattice constant $a=3$~\AA, $m_x=0.3m_0$, $m_y=2m_0$. (a) $t_x=t_y=0$~meV. (b) $t_x=20$~meV, $t_y=5$~meV. (c) $t_x=5$~meV, $t_y=20$~meV. (d) $t_x=100$~meV, $t_y=5$~meV.}
      \label{fig4}
  \end{center}
\end{figure}

Using the rectangular mBZ, we elucidate the formation of quasi-1D bands within the $\eta = 0$ M-valley. For this analysis, we fix the parameters as $\varphi = 4^\circ$, lattice constant $a = 3$~\AA, $m_x = 2m_0$ and $m_y = 0.3m_0$. We compute the moir\'e band structure across a range of interlayer tunneling strengths $t_x$ and $t_y$, presenting the results in both the conventional hexagonal and the modified rectangular mBZ, as shown in Fig.~\ref{fig4}. In the limiting case where $t_x = t_y = 0$~meV, the band structure exhibits band touching at the mBZ boundaries—an outcome of pure band folding in the absence of a moir\'e potential. Within the rectangular mBZ framework, we find that the bandwidth of the lowest-energy band is approximately $20$~meV along the $\Gamma'$–X$'$ direction and $10$~meV along the $\Gamma'$–Y$'$ direction. This anisotropy arises from the combined effects of the anisotropic effective masses and the unequal extents of the rectangular mBZ along the two momentum axes.

When setting $t_x = 20$~meV and $t_y = 5$~meV, as shown in the right column of Fig.~\ref{fig4}(b), energy gaps open at the high-symmetry points X$'$ and Y$'$, with magnitudes approximately twice the respective tunneling amplitudes—i.e., $\sim 40$~meV and $\sim 10$~meV. These gap openings effectively isolate the lowest-energy band from higher bands, yielding a topologically trivial, anisotropic band within each M-valley. The behavior becomes particularly intriguing when $t_x = 5$~meV and $t_y = 20$~meV. In this configuration, the gaps at X$'$ and Y$'$ again scale as $\sim 2t_x$ and $\sim 2t_y$, respectively. However, due to the much smaller bandwidth along the $\Gamma'$–Y$'$ direction compared to $t_y$, a nearly flat band emerges along the $y$-axis, as illustrated in Fig.~\ref{fig4}(c). This behavior closely mirrors what is observed in twisted AB-stacked Ti$_2$CO$_2$ and Zr$_2$CO$_2$ bilayers. Interestingly, anisotropic band flattening along the $x$ direction is also achievable, albeit requiring a significantly larger $t_x$ to overcome the intrinsic bandwidth along the $\Gamma'$–X$'$ direction. As demonstrated in the right column of Fig.~\ref{fig4}(d), setting $t_x = 100$~meV and $t_y = 5$~meV leads to the emergence of a flat band along the $x$-axis. This configuration corresponds to the scenario reported in twisted AA-stacked SnSe$_2$~\cite{Calugaru2024}.

From the preceding discussion, it is evident that the anisotropic band flattening observed in the lowest band of the M-valley arises from the combined effects of band folding and gap opening around the M-point. This mechanism bears a strong resemblance to the anisotropic band flattening observed in the $\Gamma$-valley moir\'e system, such as in twisted bilayer black phosphorus, as demonstrated in our previous calculations~\cite{Wang2023}. The key distinction between these two systems lies in the valley multiplicity: the M-valley features a three-fold degeneracy due to its symmetry-related counterparts, whereas the $\Gamma$-valley possesses no such degeneracy.

\emph{Discussions.} We have introduced a class of semiconducting MXenes—exemplified by Ti$_2$CO$_2$ and Zr$_2$CO$_2$—as a promising and novel platform for realizing M-valley moir\'e materials. Distinguished by the presence of three $C_{3z}$-related valleys and pronounced anisotropic band flattening, M-point moir\'e systems exhibit fundamentally different behavior compared to their $\Gamma$- and K-point twisted hexagonal counterparts. Through the construction and analysis of the corresponding moir\'e Hamiltonians, we have demonstrated that the observed anisotropic band flattening originates from band folding and gap-opening processes centered around the M-points. Notably, the direction of anisotropy is governed not by symmetry, but rather by the values of the interlayer tunneling parameters, which is different from the initial proposal in Ref.~\cite{Kariyado2019}.

This $C_{3z}$-related 2D array of quasi-1D systems may serve as a tunable platform for realizing the crossed-sliding Luttinger liquid~\cite{mukhopadhyay2001b,gao2004,du2023} and investigating non-Fermi liquid behavior. In addition, the current system resembles previously studied $p$-orbital physics in cubic and square lattices. Extending earlier exact results on itinerant ferromagnetism to this system could yield valuable insights into such phenomena~\cite{li2014,li2015}. Considerable Coulomb interactions are present in these systems, and in the presence of strong interactions and in the anisotropic limit, the combination of perfect Fermi surface nesting and the frustrated geometry of the triangular lattice may lead to a rich interplay of spin, charge, and valley density wave instabilities~\cite{zhao2008,zhang2012,Fradkin2015}. At integer fillings and within the Mott insulating regime, the underlying spin-valley exchange interactions may reveal novel spin-valley physics. For instance, the coupled-wire construction—originally developed to describe exotic 2D spin liquids~\cite{Pereira2018}—offers a promising framework for exploring such emergent spin-valley physics within our proposed platform.

\begin{acknowledgments}
\emph{Acknowledgements.} This work is supported by the Natural Science Foundation of China through Grants No.~12350404 and No.~12174066, the Innovation Program for Quantum Science and Technology through Grant No.~2021ZD0302600, the Science and Technology Commission of Shanghai Municipality under Grants No.~23JC1400600, No.~24LZ1400100 and No.~2019SHZDZX01, and is sponsored by the ``Shuguang Program'' supported by the Shanghai Education Development Foundation and Shanghai Municipal Education Commission.
\end{acknowledgments}

\emph{Note added.} During the finalization of our manuscript, we become aware of related works on a similar topic~\cite{Calugaru2024,Lei2024}. These studies also introduced M-valley moir\'e materials, but they differ from our work in terms of material proposals and theoretical analysis. Similar conclusions are reached wherever there is overlap.

\end{document}